\documentclass[a4paper]{revtex4}
\usepackage{graphicx}
\usepackage{fancyhdr}
\usepackage{amsmath}
\newcommand{\dm}{\text{\tiny DM}}

\usepackage{color}
\begin{document}
\title{Models for neutrino mass and physics beyond standard model}
\author{Amine Ahriche}
\affiliation{Department of Physics, University of Jijel, PB 98 Ouled
Aissa, DZ-18000 Jijel, Algeria\\
The Abdus Salam International Centre for Theoretical Physics, Strada
Costiera 11, I-34014, Trieste, Italy}
\author{Kristian L. McDonald}
\affiliation{ARC Centre of Excellence for Particle Physics at the
Terascale, School of Physics, The University of Sydney, NSW 2006,
Australia}
\author{Salah Nasri}
\affiliation{Physics Department, UAE University, POB 17551, Al Ain,
United Arab Emirates}

\begin{abstract}
In this work, we report on recent analysis of three-loop models of
neutrino mass with dark matter. We discuss in detail the model of
Krauss-Nasri-Trodden (KNT)~\cite{KNT}, showing that it offers a
viable solution to the neutrino mass and dark matter problems, and
describe observable experimental signals predicted by the model.
Furthermore, we show that the KNT model belongs to a larger class of
three-loop models that can differ from the KNT approach in
interesting ways.
\end{abstract}

\maketitle

\thispagestyle{fancy}

\section{Introduction}
Models with radiative neutrino mass are of significant experimental
interest. The inherent loop-suppression in such models allows the
new physics responsible for neutrino mass to be lighter than
otherwise expected. Such light new physics can be within
experimental reach, either directly, through collider experiments,
or indirectly, via e.g.~searches for lepton flavor violating (LFV)
effects. The loop suppression becomes more severe as the number of
loops increases. Thus, models with three-loop masses are
particularly interesting, as they generically require new physics at
or around the TeV scale.

He we present a class of models with radiative neutrino mass at the
three-loop level~\cite{KNT, AKS, HNOO}. We focus primarily on the
KNT model~\cite{KNT} and report recent analysis showing that the
model satisfies LFV constraints, such as $\mu \rightarrow e +
\gamma$, and fits the neutrino oscillation data. Furthermore, the
model contains a viable candidate for the dark matter (DM) in the
universe, in the form of a light right-handed (RH) neutrino. We also
show that a strongly first order electroweak phase transition can be
achieved with a Higgs mass of $\simeq 125$ GeV, as measured at the
LHC~\cite{ATLAS, CMS}. The model contains new charged scalars and we
discuss their effect on the one-loop Higgs decay to neutral gauge
bosons. Afterwards, we show that the KNT model belongs to a larger
class of related three-loop models and briefly outline their
features.


\section{The KNT Model}
The KNT model~\cite{KNT} is an extension of the SM with three RH
neutrinos, $N_{i}$, and two electrically charged scalars,
$S_{1}^{\pm}$ and $S_{2}^{\pm}$, that are singlet under the
$SU(2)_{L}$ gauge group. In addition, a discrete $Z_{2}$ symmetry is
imposed on the model, under which $\{S_{2},N_{i}\}\rightarrow
\{-S_{2},-N_{i}\}$, and all other fields are even. This symmetry
plays two key roles, preventing a tree-level coupling between $N_R$
and the SM Higgs, which would otherwise induce tree-level neutrino
masses, and ensuring that the lightest fermion $N_1$ is a stable DM
candidate. The Lagrangian reads
\begin{align}
\mathcal{L} &
=\mathcal{L}_{SM}+\{f_{\alpha\beta}L_{\alpha}^{T}Ci\tau
_{2}L_{\beta}S_{1}^{+}+g_{i\alpha}N_{i}S_{2}^{+}\ell_{\alpha R}
+\tfrac{1}{2}m_{N_{i}}N_{i}^{C}N_{i}+h.c\}-V(\Phi,S_{1},S_{2}),
\label{L}
\end{align}
where $L_{\alpha}$ is the left-handed lepton doublet, $f_{\alpha\beta}$ are
Yukawa couplings which are antisymmetric in the generation indices $\alpha$
and $\beta$, $m_{N_{i}}$\ are the Majorana RH neutrino masses, $C$ is the
charge conjugation matrix, and $V(\Phi,S_{1},S_{2})$ is the tree-level scalar
potential, which is given by
\begin{align}
V(\Phi,S_{1,2}) & =\lambda\left( \left\vert \Phi\right\vert
^{2}\right)
^{2}-\mu^{2}\left\vert \Phi\right\vert ^{2}+m_{1}^{2}S_{1}^{\ast}S_{1}%
+m_{2}^{2}S_{2}^{\ast}S_{2}+\lambda_{1}S_{1}^{\ast}S_{1}\left\vert
\Phi\right\vert ^{2}+\lambda_{2}S_{2}^{\ast}S_{2}\left\vert \Phi\right\vert
^{2}\nonumber\\
& +\frac{\eta_{1}}{2}\left( S_{1}^{\ast}S_{1}\right) ^{2}+\frac{\eta_{2}%
}{2}\left( S_{2}^{\ast}S_{2}\right) ^{2}+\eta_{12}S_{1}^{\ast}S_{1}%
S_{2}^{\ast}S_{2}+\left\{
\lambda_{s}S_{1}S_{1}S_{2}^{\ast}S_{2}^{\ast }+h.c\right\} .
\label{nu-mass}
\end{align}
Here $\Phi$ denotes the SM Higgs doublet.

\subsection{Neutrino Mass and LFV}

\begin{figure}[t]
\begin{centering}
\includegraphics[width=6cm,height=2.2cm]{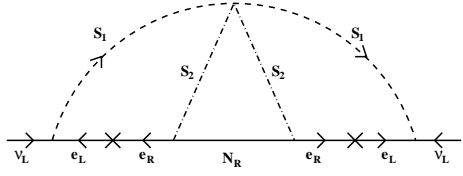}
\par\end{centering}
\caption{\textit{The three-loop diagram that generates the neutrino
mass.}} \label{diag}
\end{figure}

The neutrino mass matrix elements, arising from the three-loop diagram in Fig.
\ref{diag}, are given by
\begin{equation}
(M_{\nu})_{\alpha\beta}=\frac{\lambda_{s}m_{\ell_{i}}m_{\ell_{k}}}{\left(
4\pi^{2}\right) ^{3}m_{S_{2}}}f_{\alpha i}f_{\beta
k}g_{ij}g_{kj}F\left(
m_{N_{j}}^{2}/m_{S_{2}}^{2},m_{S_{1}}^{2}/m_{S_{2}}^{2}\right)
,\label{nu-mass-1}%
\end{equation}
where $\rho,\kappa(=e,\mu,\tau)$ are the charged leptons flavor
indices, $i=1,2,3$ denotes the three RH neutrinos, and the function
$F$ is a loop integral which is $\mathcal{O}(1)$~\cite{AN2013}. It
is interesting to note that, unlike the conventional seesaw
mechanism, the radiatively generated neutrino masses are directly
proportional to the charged lepton and RH neutrino masses, as well
as being loop-suppressed.

The Lagrangian (\ref{L}) induces flavor violating processes such as
$\ell_{\alpha}\rightarrow\gamma\ell_{\beta}$ for
$m_{\ell_{\alpha}}>m_{\ell_{\beta}}$, generated at one loop via the
exchange of the charged scalars $S_{1,2}^{\pm}$. The
branching ratio of such process is given by
\begin{align}
B(\ell_{\alpha} &
\rightarrow\gamma\ell_{\beta})=\frac{\alpha_{em}\upsilon^{4}}{384\pi}\left\{
\frac{\left\vert
f_{\kappa\alpha}f_{\kappa\beta}^{\ast}\right\vert ^{2}}{m_{S_{1}}^{4}}%
+\frac{36}{m_{S_{2}}^{4}}\left\vert
{\sum\limits_{i}}g_{i\alpha}g_{i\beta }^{\ast}F_{2}\left(
\frac{m_{N_{i}}^{2}}{m_{S_{2}}^{2}}\right) \right\vert ^{2}\right\}
,\label{mutoegamma}
\end{align}
with $\kappa\neq\alpha,\beta$, $\alpha_{em}$ being the fine structure constant
and $F_{2}(x)=(1-6x+3x^{2}+2x^{3}-6x^{2}\ln x)/6(1-x)^{4}$. For the case of
$\ell_{\alpha}=\ell_{\beta}=\mu$, this leads to a new contribution to the muon
anomalous magnetic moment, $\delta a_{\mu}$.\\

In our scan of the parameter space of the model we impose the
experimental bound $\mathbf{B}\left( \mathbf{\mu\rightarrow
e\gamma}\right) < 5.7\times10^{-13}$~\cite{LFV}; and use the allowed
values for the neutrino mixing $s_{12}^{2}=0.320_{-0.017}
^{+0.016}$, $s_{23}^{2}=0.43_{-0.03}^{+0.03}$,
$s_{13}^{2}=0.025_{-0.003}^{+0.003}$, and the mass squared
difference $\left\vert \Delta m_{31}^{2}\right\vert
=2.55_{-0.09}^{+0.06}\times10^{-3}$ \textrm{eV}$^{2}$ and $\Delta
m_{21}^{2}=7.62_{-0.19}^{+0.19}\times10^{-5}\mathrm{eV}^{2}$
~\cite{GF}.

\subsection{Dark matter}

An immediate implication of the $Z_{2}$ symmetry is that that
lightest right handed neutrino, $N_{1}$, is stable, and hence a
candidate for dark matter (DM). The $N_{1}$ number density get
depleted through the process
$N_{1}N_{1}\rightarrow\ell_{\alpha}\ell_{\beta}$ via the $t$ and u
channels exchange of $S_{2}^{\pm}$. In the non-relativistic limit,
the total annihilation cross section is given by
\begin{equation}
\sigma_{N_{1}N_{1}}\upsilon_{r}\simeq\sum_{\alpha,\beta}|g_{1\alpha}g_{1\beta
}^{\ast}|^{2}\frac{m_{N_{1}}^{2}\left(
m_{S_{2}}^{4}+m_{N_{1}}^{4}\right)
}{48\pi\left( m_{S_{2}}^{2}+m_{N_{1}}^{2}\right) ^{4}}\upsilon_{r}%
^{2},\label{sig11}%
\end{equation}
with $\upsilon_{r}$ is the relative velocity between the annihilation $N_{1}%
$'s. The relic density after the decoupling of $N_{1}$ can be
obtained by solving the Boltzmann equation, and it is approximately
given by

\begin{align}
\Omega_{N_{1}}h^{2} &
\simeq\frac{1.28\times10^{-2}}{\sum_{\alpha,\beta}|g_{1\alpha}g_{1\beta
}^{\ast}|^{2}}\left( \frac{m_{N_{1}}}{135~\mathrm{GeV}}\right) ^{2}%
\frac{\left( 1+m_{S_{2}}^{2}/m_{N_{1}}^{2}\right) ^{4}}{1+m_{S_{2}}%
^{4}/m_{N_{1}}^{4}},\label{Omh2}
\end{align}
where ${<\upsilon_{r}^{2}>}\simeq6/x_{f}\simeq6/25$ is the thermal
average of the relative velocity squared of a pair of two $N_{1}$
particles, $M_{pl}$ is planck mass; and $g_{\ast}(T_{f})$ is the
total number of effective massless degrees of freedom at the
freeze-out temperature $T_{f} \sim m_{N_{1}}/{25}$\textbf{.}

\begin{figure}[t]
\label{Om}\begin{centering}
\includegraphics[width=7.5cm,height=5.5cm]{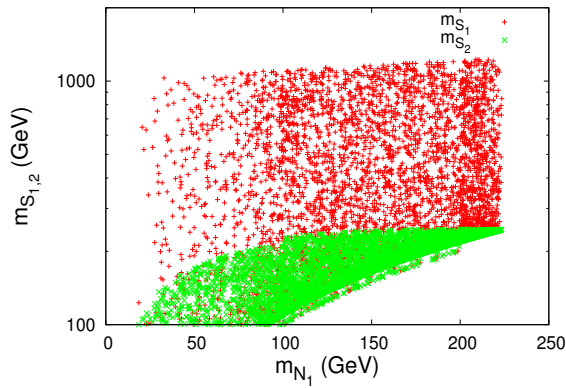}
\par\end{centering}
\caption{\textit{The charged scalar masses $m_{S_{1}}$ (red) and $m_{S_{2}}$
(green) versus the lightest RH neutrino mass, where the consistency with the
neutrino data, LFV constraints and the DM relic density have been imposed.}}%
\label{msmn}%
\end{figure}

In Fig. \ref{msmn}, we plot the allowed mass range for
$(m_{N_{1}},m_{S_{i}})$ that gives the observed dark matter relic
density ~\cite{Planck}. As seen in the figure, the neutrino
experimental data, the bound on LFV process, combined with the relic
density, seems to prefer $m_{S_{1}}>m_{S_{2}}$ for large regions of
parameter space. However, the masses of both the DM and the charged
scalar $S_{2}^{\pm}$ are bounded from above, with $m_{N_{1}}<225$
\textrm{GeV}\ and $m_{S_{2}}<245$ \textrm{GeV},
respectively.\textbf{\ }

\subsection{Electroweak Phase Transition}

Although the SM has all the qualitative ingredients for electroweak
baryogenesis, the amount of matter-antimatter asymmetry generated is
too small. One of the reasons has to with the fact that the
electroweak phase transition (EWPT) is not strongly first order,
which is necessary to suppress the sphaleron processes in the broken
phase. The strength of the EWPT can be improved if there are new
scalar degrees of freedom around the electroweak scale coupled to
the SM Higgs, which is the case in the KNT model.

\begin{figure}[h]
\includegraphics[width=7.5cm,height=5.5cm]{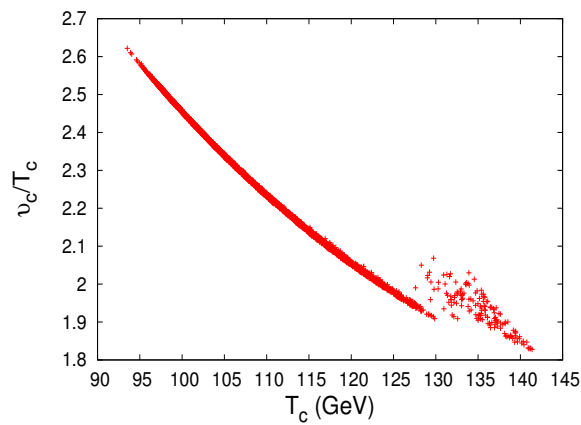}
\caption{\textit{The critical temperature is presented versus the
quantity $\upsilon_{c}/T_{c}$.}}
\label{ct}%
\end{figure}

The investigation of the scalar effective potential reveals that,
within the allowed parameter space of the model, the strength of the
electroweak phase transition (EWPT) can be first
order~\cite{AN2013}. We find that if the one-loop corrections to the
Higgs mass are sizeable, then the condition
$\upsilon(T_{c})/T_{c}>1$ for having a first order EWPT can
 be realized while keeping the Higgs mass around
$125$~GeV. The reason for this being that the extra charged singlets
affect the dynamics of the SM scalar field VEV around the critical
temperature~\cite{hna}. In Fig. \ref{ct}, we show the plot for
$\upsilon(T_{c})/T_{c}$ versus the critical temperature and one
observes that a strongly first order EWPT is possible while the
critical temperature lies around 100 $\mathrm{GeV}$.

\subsection{KNT at High Energy Lepton colliders}

\begin{figure}[ptb]
\begin{centering}
\includegraphics[width=7.5cm,height=5.5cm]{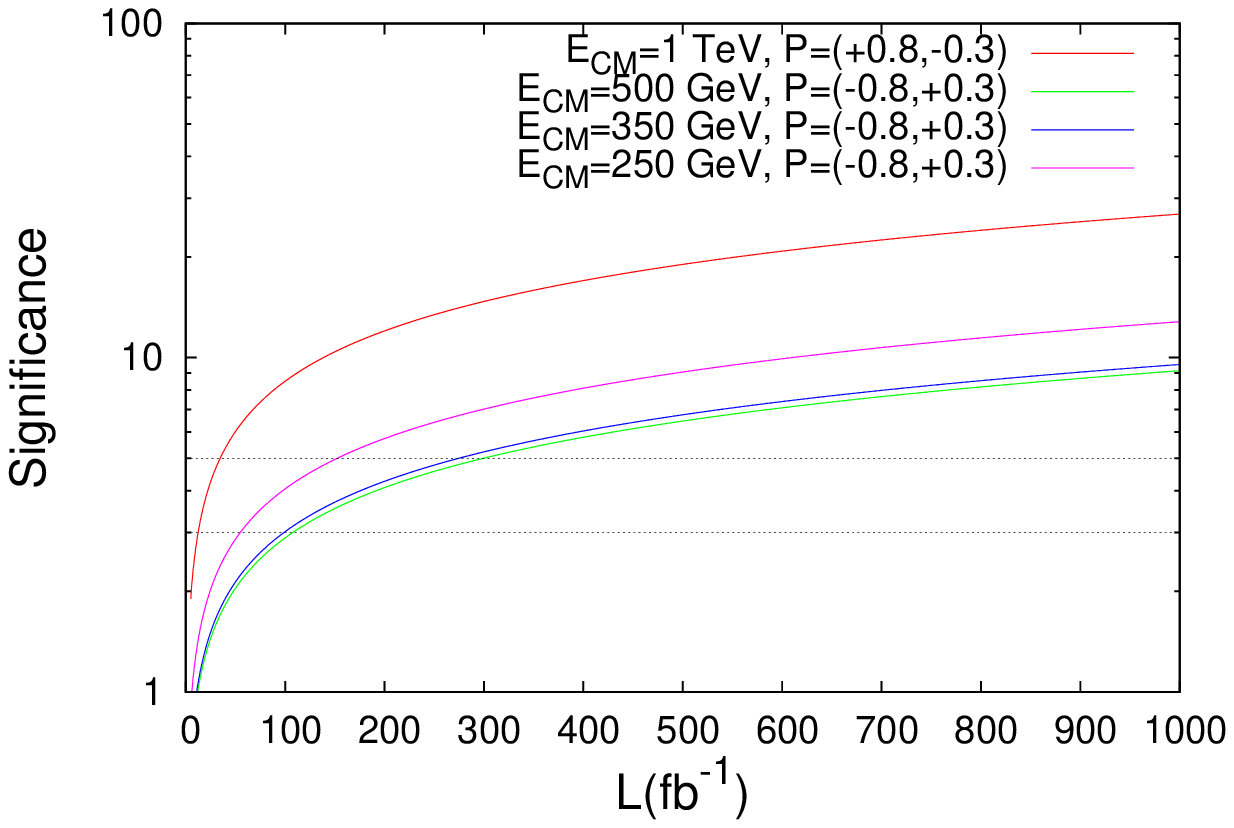}\includegraphics[width=7.5cm,height=5.5cm]{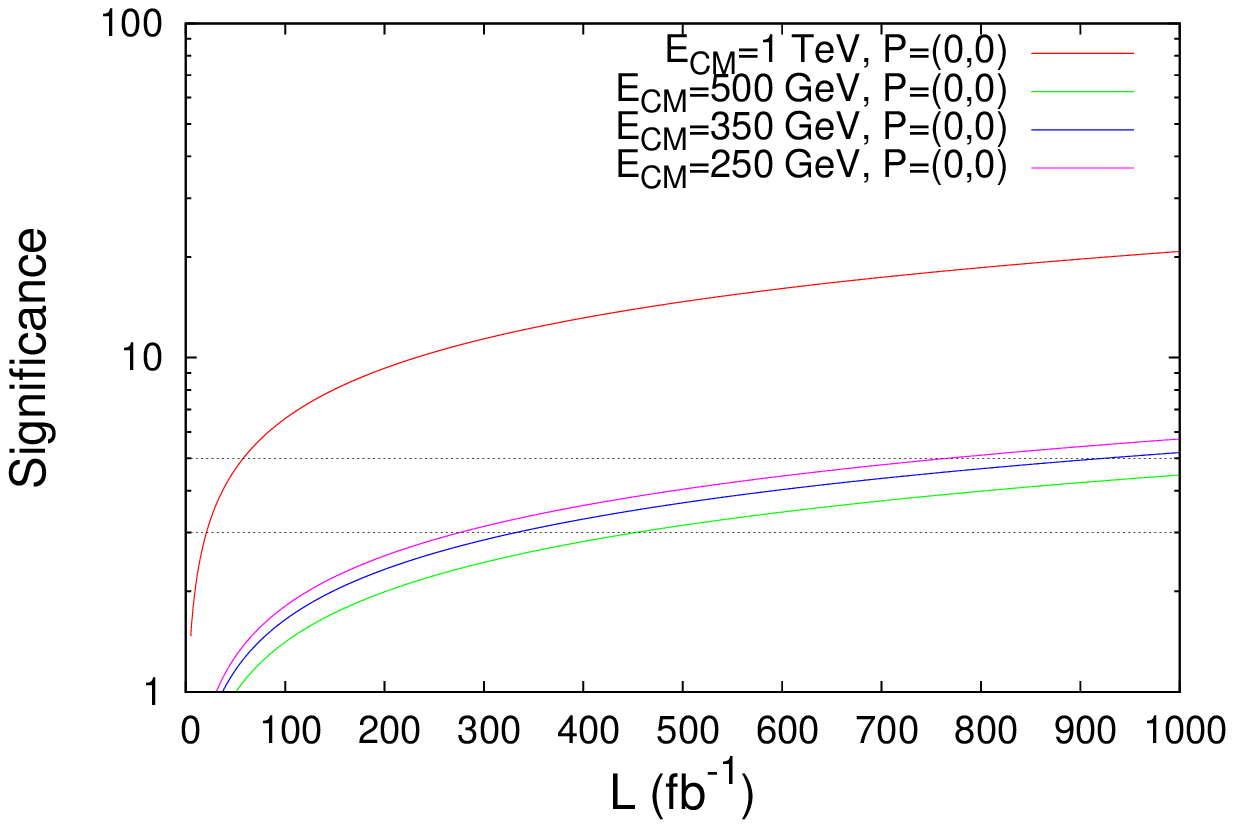}
\par\end{centering}
\caption{\textit{The significance versus luminosity at different CM
energies within the cuts defined in Table-\protect\ref{T1}; with
(left) and without (right) polarized beams. The two horizontal
dashed lines represent $\mathcal{S}=3$ and $\mathcal{S}=5$,
respectively.}} \label{SvsL}
\end{figure}

Since the RH neutrinos couple to the charged leptons, one excepts
them to be produced at $e^{-}e^{+}$ colliders, such as the ILC and
CLIC, with a collision energy $\sqrt{s}$ of few hundreds of
\textrm{GeV} up to a TeV. If the produced pairs are the form
$N_{2,3}N_{2,3}$ or $N_{1}N_{2,3}$, then $N_{2,3} $ will decay into
a charged lepton and $S_{2}^{\pm}$, and subsequently $S_{2}^{\pm}$
decays into $N_{1}$ and a charged lepton. In addition, the SM
neutrinos will be also produced via the decay of the charged scalar
$S_1$. Here we concentrate on he process $e^{-}e^{+}\rightarrow
e^{-}\mu^{+}+E_{miss}$~\cite{ANS}, where the missing energy
corresponds to any state in the set $\mathcal{E}_{miss}\subset \{\nu
_{\mu }\bar{\nu}_{e}$, $\nu _{e}\bar{\nu}_{\tau }$, $\nu _{\tau
}\bar{\nu}_{e}$, $\nu _{\mu }\bar{\nu}_{\mu }
$, $\nu _{\tau }\bar{\nu}_{\mu }$, $\nu _{\tau }\bar{\nu}_{\tau }$, $%
N_{i}N_{k};~i,k=1,2,3\}$. The total expected cross section of the processes $%
e^{-}e^{+}\rightarrow e^{-}\mu ^{+}+E_{miss}$ is represented by $\sigma
^{EX}$, while $\sigma (\mathcal{E}_{miss})$ denotes the cross section of
different subprocesses. As a benchmark, we consider
\begin{align}
f_{e\mu }& =-(4.97+i1.41)\times 10^{-2},~f_{e\tau }=0.106+i0.0859,~f_{\mu
\tau }=(3.04-i4.72)\times 10^{-6}, \notag \\
g_{i\alpha }& =10^{-2}\times \left(
\begin{array}{ccc}
0.2249+i0.3252 & 0.0053+i0.7789 & 0.4709+i1.47 \\
1.099+i1.511 & -1.365-i1.003 & 0.6532-i0.1845 \\
122.1+i178.4 & -0.6398-i0.6656 & -10.56+i68.56%
\end{array}%
\right) , \notag \\
m_{N_{i}}& =\{162.2\mathrm{~GeV},\ 182.1\mathrm{~GeV},~209.8\mathrm{~GeV}%
\},~m_{S_{i}}=\{914.2\mathrm{~GeV},~239.7\mathrm{~GeV}\}, \label{bA}
\end{align}
We use LanHep and CalcHep to simulate our model and generate the
differential cross section and the relevant kinematic variables for
different CM energy: $E_{CM}=$250 , 350, 500 \textrm{GeV} and1
\textrm{TeV}, with unpolarized beams at first; and then we consider
the polarized beams with $P\left( e^{-},e^{+}\right) =[-0.8,+0.3]$
and/or $P\left( e^{-},e^{+}\right) =[+0.8,-0.3]$. Imposing the
appropriate cuts given in Table-\ref{T1}, we show in Fig.~\ref{SvsL}
the dependance of the significance on the accumulated luminosity
with and without polarized beams for the considered CM energies. We
clearly see that for a polarized beam, the signal can be observed
even with relatively low integrated luminosity. For example, at
$E_{CM}=250$ \textrm{GeV}, the 5 $\sigma$ required luminosity is 150
$fb^{-1}$ for polarized beam as compared to 700 $fb^{-1}$ without
polarization.

\begin{table}[ptb]
\begin{tabular}{|c|c|}
\hline\hline
$E_{CM}(GeV)$ & Selection cuts \\ \hline
$250$ & $%
\begin{array}{c}
70~GeV<E_{\ell}<110~GeV,~70~GeV<M_{e,\mu}<220~GeV, \\
M_{miss}<120~GeV,~0.5<\eta_{e}<2,~-2<\eta_{\mu}<-0.5.%
\end{array}
$ \\ \hline
$350$ & $%
\begin{array}{c}
90~GeV<E_{\ell}<165~GeV,~100~GeV<M_{e,\mu}<280~GeV, \\
M_{miss}<200~GeV,~0.5<\eta_{e}<3,~-2.5<\eta_{\mu}<0.%
\end{array}
$ \\ \hline
$500$ & $%
\begin{array}{c}
120~GeV<E_{\ell}<240~GeV,~300~GeV<M_{e,\mu}<480~GeV, \\
M_{miss}<300~GeV,~0.5<\eta_{e}<3,~-3<\eta_{\mu}<0.%
\end{array}
$ \\ \hline
$1000$ & $%
\begin{array}{c}
E_{\ell}<70~GeV,~M_{e,\mu}<140~GeV,~M_{miss}>750~GeV \\
,~0.1<\eta_{e}<0.8,~-0.8<\eta_{\mu}<-0.1.%
\end{array}
$ \\ \hline\hline
\end{tabular}
\caption{\textit{Relevant cuts for the process $e^{+}e^{-}\rightarrow
E_{miss}+e^{-}\protect\mu^{+}$ for different CM energies. Here $E_{\ell}$
and $\protect\eta_{\ell}$ are the charged lepton energy in its
pseudo-rapidity, $M_{\mathit{e,}\protect\mu}$ is the electron-muon invariant
mass and $M_{miss}$ is the missing invariant mass.}}
\label{T1}
\end{table}

\section{A Class of Three-Loop Models with Dark Matter}

The KNT model is a simple theory that is capable of simultaneously
addressing the neutrino mass and DM problems. It predicts new
physics that can be probed at colliders and precision experiments.
Within the KNT model, the DM is part of a $Z_2$-odd sector that
propagates in the inner loop of the neutrino mass diagram (see
Figure~\ref{diag}). One might wonder whether generalizations of the
KNT model exist. It is well known that the seesaw mechanism can be
generalized to a type-III variant with $SU(2)$ triplet
fermions~\cite{Foot:1988aq}, and a further variant with quintuplet
fermions~\cite{Kumericki:2012bh}. It is perhaps not surprising,
therefore, to learn that the KNT model forms part of a larger class
of theories that similarly achieve neutrino mass at the three-loop
level, with DM propagating in the inner loop.

The generalized models require that the SM be extended to include
the charged singlet scalar $S_1^+\sim(1,1,2)$, a new scalar $S_2$,
and three real fermions $N_R$. However, instead of taking $S_2$ and
$N_R$ as $SU(2)_L$ singlets, the variant models employ larger
multiplets. The resulting models share key similarities with the KNT
model, with the neutrino mass diagram retaining the exact form in
Figure~\ref{diag}, and the DM remaining as the lightest fermion
$N_R$. However, there are also some interesting differences,
relative to the KNT model. Firstly, the DM is now part a larger
multiplet with non-trivial gauge interactions. This modifies
expectations for the DM mass and allows new DM interactions.
Secondly, there are interesting consequences for the $Z_2$ symmetry,
which depend on the specific details of the model, as we now
outline:
\begin{itemize}

\item \underline{The triplet model}: Here $S_2\sim(1,3,2)$ and $N_R\sim(1,3,0)$ are taken as $SU(2)_L$ triplets~\cite{Ahriche:2014cda}. Similar to the KNT model, the triplet model requires a $Z_2$ symmetry with action $\{S_{2},N_{R}\}\rightarrow \{-S_{2},-N_{R}\}$ to prevent tree-level neutrino masses and ensure a stable DM candidate. The DM is the neutral component of the lightest triplet fermion and has extra Yukawa interactions mediated by $g_{i\alpha}\ne0$ in Eq.~\eqref{L}. Consequently its mass increases to $M_\dm\approx
3$~TeV~\cite{Ahriche:2014cda}.

\item \underline{The quintuplet model}: This model employs the multiplets $S_2\sim(1,5,2)$ and $N_R\sim(1,5,0)$~\cite{Ahriche:2014oda}. Unlike the KNT model and the triplet model, the quintuplet variant does not require a $Z_2$ symmetry to prevent tree-level neutrino mass. It is therefore a viable model of radiative neutrino mass, independent of DM considerations. Interestingly, the $Z_2$ symmetry $\{S_{2},N_{R}\}\rightarrow \{-S_{2},-N_{R}\}$ is broken by a single coupling $\lambda$ in the model; for technically-natural values of $\lambda\ll1$, one obtains a long-lived DM candidate, while in the limit $\lambda\rightarrow0$ the $Z_2$ symmetry becomes exact and the DM is absolutely stable~\cite{Ahriche:2014oda}. The quintuplet model is therefore a model of radiative neutrino mass, with or without DM, depending on the region of parameter space considered. Due to the gauge interactions, the case with DM requires $M_\dm\approx10$~TeV.

\item \underline{The septuplet model}: This case has $S_2\sim(1,7,2)$ and $N_R\sim(1,7,0)$, and is of particular interest~\cite{Ahriche:2015wha}. Similar to the quintuplet model, the $Z_2$ symmetry is not required to prevent tree-level neutrino mass. However, different from the quintuplet model, the septuplet model \emph{automatically} possesses an
accidental $Z_2$ symmetry with action $\{S_{2},N_{R}\}\rightarrow
\{-S_{2},-N_{R}\}$, and therefore always contains a stable DM
candidate. This gives a common description for neutrino mass and DM
without requiring any new symmetries. The DM is relatively heavy,
with $M_\dm\approx20-25$~TeV, due to the Sommerfeld enhancement
induced by $SU(2)$ gauge boson exchange~\cite{Ahriche:2015wha}.

\end{itemize}
Detailed studies show that the variant models have large regions of
viable parameter space that fit the neutrino oscillation data,
reproduce the observed DM relic density, and satisfy experimental
constraints from, e.g, LFV
searches~\cite{Ahriche:2014cda,Ahriche:2014oda,Ahriche:2015wha}. In
each of the variant models, the DM is a heavy Majorana fermion and
the mass for $S_2$ should obey $M_2>M_\dm$, placing both multiplets
beyond the reach of current colliders. However, the scalar $S_1$ can
retain an $\mathcal{O}(100)$~GeV mass in all cases, offering the
best chance for testing the models at colliders. The models with
larger $SU(2)_L$ multiplets also generate sizable contributions to
LFV processes like $\mu\rightarrow e+\gamma$. In fact, these are
typically enhanced relative to the KNT case, improving the prospects
for testing the model in future LFV
searches~\cite{Ahriche:2014cda,Ahriche:2014oda,Ahriche:2015wha}. Due
to the new gauge interactions for the DM, direct-detection prospects
also improve for the models with larger
multiplets~\cite{Ahriche:2014cda,Ahriche:2014oda,Ahriche:2015wha}.

\begin{figure}[t]
\begin{centering}
 \includegraphics[width=7cm,height=3cm]{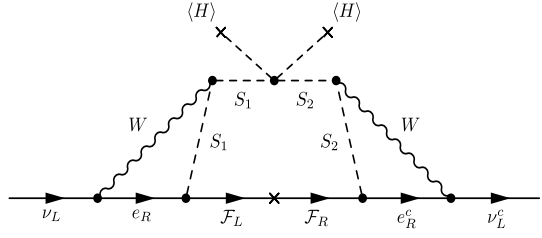}
\par\end{centering}
\caption{\textit{Three-loop diagram for neutrino mass in a variant
model with $S_1\sim(1,2,1)$, $S_2\sim(1,2,3)$ and $F\sim(1,2,-1)$.
The beyond-SM multiplets $S_{1,2}$ and $F$ are taken odd under a
$Z_2$ symmetry and the dark matter belongs to the inert doublet
$S_1$.}} \label{general}
\end{figure}

There are additional models of neutrino mass with DM that are
related to the KNT model~\cite{Chen:2014ska}. For example, one can
replace the internal SM leptons in Figure~\ref{diag} with down-type
quarks, $e_{L,R}\rightarrow d_{L,R}$, and take
$S_1\sim(\bar{3},1,2/3)$ and $S_2\sim(3,1,-2/3)$ as colored
multiplets. Further colored variants are also
possible~\cite{Chen:2014ska}. Alternatively, one may replace the
real (Majorana) fermion $N_R$ with a complex (Dirac) fermion and
consider a modified topology for the loop-diagram, with the DM being
an inert scalar (see Figure~\ref{general} for an example). A
systematic description of these variant models appears in
Ref.~\cite{Chen:2014ska}.

\section{Conclusion}

Models with radiative neutrino mass offer a promising way to
experimentally discern the new physics responsible for the origin of
neutrino mass. Three-loop models are particularly interesting as the
new physics is generically expected at or around the TeV scale.
Furthermore, such models can provide viable DM candidates, thereby
solving both the neutrino mass and DM problems. Here we reported
recent analysis of the KNT model, which show the model to be a
viable theory for neutrino mass and DM that can be tested at
collider experiments. We also showed that the KNT model belongs to a
larger class of three-loop models that can generate similarly
interesting observable effects.


\begin{acknowledgments}
S. Nasri would like to thank Shinya Kanemura and all the members of
the organizing committee of HPNP2015 for the very pleasing meeting
and their great hospitality in Toyama. The authors thank R. Soualah,
C.~S.~Chen and T.~Toma for productive collaborations in this field.
\end{acknowledgments}


\begin{thebibliography}{99}
\bibitem{KNT} L.~M.~Krauss, S.~Nasri and M.~Trodden,
 Phys.\ Rev.\ D {\bf 67}, 085002 (2003).
 \bibitem{AKS} M.~Aoki, S.~Kanemura and O.~Seto,
 Phys.\ Rev.\ Lett.\ {\bf 102}, 051805 (2009).
M.~Gustafsson, J.~M.~No and M.~A.~Rivera,
 Phys.\ Rev.\ Lett.\ {\bf 110}, no. 21, 211802 (2013)
 [Erratum-ibid. 112, no. 25, 259902 (2014)].
\bibitem{HNOO} H.~Hatanaka, K.~Nishiwaki, H.~Okada and Y.~Orikasa,
 Nucl.\ Phys.\ B {\bf 894}, 268 (2015).
\bibitem {ATLAS}G. Aad et al. (ATLAS Collaboration), Phys. Lett. B 716, 1-29 (2012).
\bibitem{CMS} S. Chatrchyan et al. (CMS Collaboration), Phys. Lett. B 716, 30-61 (2012).
 \bibitem{AN2013} A.~Ahriche and S.~Nasri,
 JCAP {\bf 1307}, 035 (2013).
\bibitem {LFV}J.~Adam {\it et al.}  [MEG Collaboration],
  Phys.\ Rev.\ Lett.\  {\bf 110}, 201801 (2013)
  [arXiv:1303.0754 [hep-ex]].
\bibitem {GF}D.V. Forero, M. Tortola and J.W.F. Valle, Phys. Rev. D 86, 073012 (2012).

\bibitem {Planck}P.A.R. Ade et al. [Planck Collaboration], Astron.\ Astrophys.\ {\bf 571}, A1 (2014).

\bibitem{hna} A. Ahriche, Phys. Rev. D75, 083522 (2007); A. Ahriche and S.
Nasri, Phys. Rev. D 83, 045032 (2011); Phys. Rev. D 85, 093007
(2012).

\bibitem{ANS} A. Ahriche, S. Nasri and R. Soualah, Phys. Rev. D 89, 095010 (2014).

\bibitem{Foot:1988aq}
 R.~Foot, H.~Lew, X.~G.~He and G.~C.~Joshi,
 Z.\ Phys.\ C {\bf 44}, 441 (1989).
\bibitem{Kumericki:2012bh}
 K.~Kumericki, I.~Picek and B.~Radovcic,
 Phys.\ Rev.\ D {\bf 86}, 013006 (2012)
 [arXiv:1204.6599 [hep-ph]];
 Y.~Liao,
 JHEP {\bf 1106}, 098 (2011)
 [arXiv:1011.3633 [hep-ph]].
\bibitem{Ahriche:2014cda}
 A.~Ahriche, C.~S.~Chen, K.~L.~McDonald and S.~Nasri,
 Phys.\ Rev.\ D {\bf 90}, 015024 (2014)
 [arXiv:1404.2696 [hep-ph]].
\bibitem{Ahriche:2014oda}
 A.~Ahriche, K.~L.~McDonald and S.~Nasri,
 JHEP {\bf 1410}, 167 (2014)
 [arXiv:1404.5917 [hep-ph]].
\bibitem{Ahriche:2015wha}
 A.~Ahriche, K.~L.~McDonald, S.~Nasri and T.~Toma,
 arXiv:1504.05755 [hep-ph].
\bibitem{Chen:2014ska}
 C.~S.~Chen, K.~L.~McDonald and S.~Nasri,
 Phys.\ Lett.\ B {\bf 734}, 388 (2014)
 [arXiv:1404.6033 [hep-ph]].

\end{thebibliography}
\end{document}